# Quantum fluctuations associated with first-order magnetic transition in a frustrated kagome lattice antiferromagnet

Zhongchen Xu,[1,2,3,#] Xinyang Liu,[1,6,#] Cuiwei Zhang,[1,2] Shuai Zhang,[1,4] Feng Jin,[1] Junsen Xiang,[1,4] Quansheng Wu,[1,4,5] Xianmin Zhang,[3,*] Peijie Sun,[1,4,5,*] Youguo Shi[1,2,5,*]

[1]*Beijing National Laboratory for Condensed Matter Physics and Institute of Physics, Chinese Academy of Sciences, Beijing 100190, China*

[2]*Center of Materials Science and Optoelectronics Engineering, University of Chinese Academy of Sciences, Beijing 100190, China*

[3]*Key Laboratory for Anisotropy and Texture of Materials (Ministry of Education), School of Material Science and Engineering, Northeastern University, Shenyang 110819, China*

[4]*School of Physical Sciences, University of Chinese Academy of Sciences, Beijing 100190, China*

[5]*Songshan Lake Materials Laboratory, Dongguan, Guangdong 523808, China*

[6]*School of Physics, Beihang University, Beijing 100191, China*

Correspondence: Xianmin Zhang (zhangxm@atm.neu.edu.cn), Peijie Sun (pjsun@iphy.ac.cn) and Youguo Shi (ygshi@iphy.ac.cn)

Zhongchen Xu and Xinyang Liu contributed equally to this work

## Abstract

Intense quantum fluctuations arising from geometrical frustrations in kagome-lattice magnets provide a feasible approach to exotic quantum states. Here, we document an unexpected isosymmetric first-order magnetic transition in the recently synthesized frustrated kagome-lattice antiferromagnet $Nd_3ScBi_5$, which is characterized by significant latent heat and a pronounced magnetocaloric effect, as well as discontinuous Raman shifts and negligible hysteresis. Employing the magnetocaloric effect as a detection method, in conjunction with systematical field-dependent physical properties, we uncover a distinctive 1/2 magnetization plateau phase with significant quantum fluctuations. Our study unveils $Nd_3ScBi_5$ as a prototypical model with an emerging phase of enhanced quantum fluctuations triggered by first-order magnetic transitions.

## Introduction

First-order magnetic transitions (FOMT), characterized by abrupt changes in state functions such as free energy is often intricately interwoven with the crystal structure. Magnetic material holds considerable technological promise, particularly in the realms of magnetic shape memory[1] and magnetostriction[2], as well as emerging magnetocaloric effects (MCE) for solid-state refrigeration[3, 4]. Moreover, the highly frustrated kagome lattice magnet serves as a fertile ground for the emergence of exotic quantum spin states. For instance, the two-dimensional frustrated antiferromagnet $Zn_xCu_{4-x}(OD)_6Cl_2$ experimentally exhibits characteristics of quantum spin singlet states in solids, thereby breaking translational symmetry[5]. The distorted kagome intermetallic compound CePdAl gives rise to a paramagnetic quantum-critical phase due to the competitive interplay of geometric frustrations, Ruderman-Kittel-Kasuya-Yosida interactions, and Kondo effects[6]. The kagome metal $Ni_3In$ demonstrates strange metallic behavior and quantum criticality, with the localized state originating from band flattening induced by destructive interference among partially filled bands[7].

The $Ln_3MX_5$ family ($Ln$ = rare earth elements, $M$ = transition materials, and $X$ = As, Sb, or Bi) has come into our research purview, that hosts a distorted kagome motif and serves as an excellent carrier for the study of geometrical frustration frameworks[8, 9]. Previous studies have systematically explored fractional magnetization platforms and singular anisotropic transport in the frustrated Kondo-lattice compound $Ce_3ScBi_5$[10]. One intriguing conjecture is to design an environment where the prominent Kondo effect is absent, allowing for a purer investigation of spin-frustrated interactions within the $Ln_3MX_5$ family.

In this work, $Nd_3ScBi_5$ single crystal with anti-$Hf_5Sn_3Cu$ hexagonal structure was successfully synthesized via the bismuth self-flux method, leading to the formation of $Nd^{3+}$ arranged in a frustrated structure within a distorted kagome-lattice. We elucidate an intriguing mechanism driving a first-order antiferromagnetic (AFM) transition characterized by substantial latent heat and discontinuous Raman shifts, alongside minimal magnetic and thermal hysteresis. Remarkably, the MCE associated with strong quantum fluctuations manifests as distinct valley-like regimes during demagnetization cooling. The identification of novel quantum criticality and fractional magnetization phases in this ideal FOMT material opens new avenues for exploring exotic quantum phenomena.

## Results

### Frustrated kagome-lattice antiferromagnets

$Nd_3ScBi_5$ is a novel compound distinguished by its quasi-one-dimensional (1D) structural motif, crystallizing within the $P6_3/mcm$ space group. This crystal structure incorporates three quasi-1D substructures: hypervalent $Bi^{2-}$ chains, face-sharing $ScBi_6$ octahedra, and zig-zag $Nd^{3+}$ chains, all aligned along the c-axis. Notably, the *6g* Nd sites within the *ab*-plane form a geometrically frustrated, distorted kagome-lattice and are stacked along the crystallographic $2_1$ screw axis (Fig. 1a). This distortion originates from the counter-rotational twist of the triangular lattice within the kagome crystal around the *c*-axis, thereby disrupting the spatial inversion symmetry of individual kagome layers[11]. Supplementary Notes 1 and 2 provide further details on the crystal structure and its characterization.

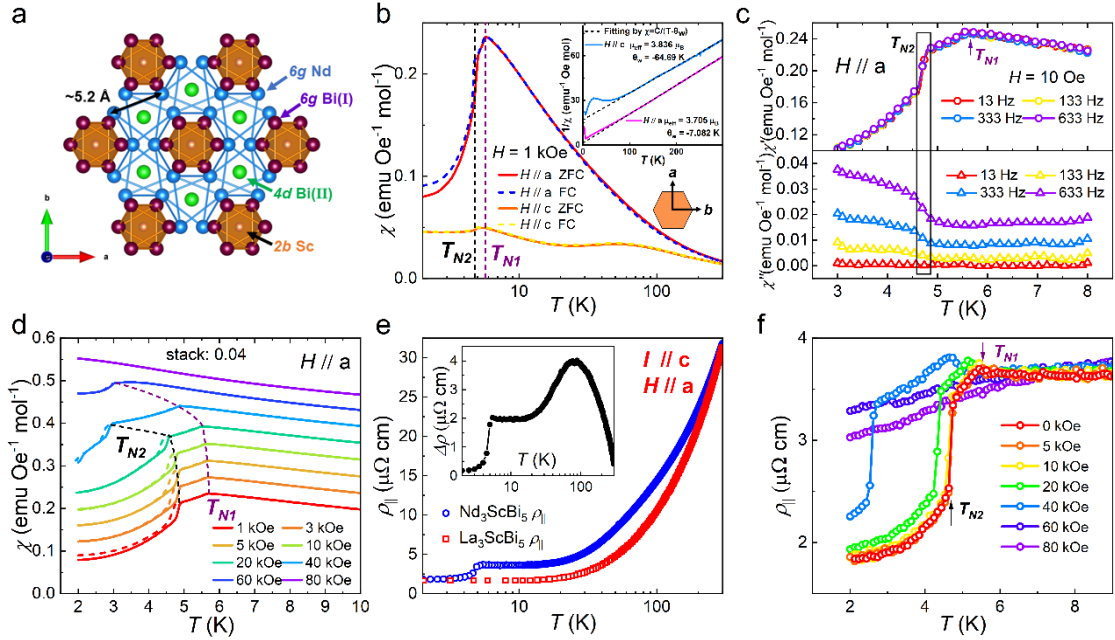

**Fig. 1. Crystal structure, Magnetic susceptibility ($\chi$), and Resistivity ($\rho$) of Nd$_3$ScBi$_5$.** (a) Crystal structure for Nd$_3$ScBi$_5$ viewed along the c-axis, illustrating the distorted kagome-lattice formed by 6g Nd atoms and the crystallographic $2_1$ screw axis. (b) Temperature-dependent magnetic susceptibility and inverse magnetic susceptibility of Nd$_3$ScBi$_5$ for *H // a* and *H // c* (*H* = 1 kOe). (c) Real ($\chi'$) and imaginary ($\chi''$) components of the AC susceptibility measured with an external magnetic field of 10 Oe at various frequencies. (d) Temperature-dependent magnetic susceptibility under different magnetic fields for *H // a*, with zero-field-cooled (ZFC) and field-cooled (FC) data shown by solid and dashed lines, respectively. (e) Temperature-dependent electrical resistivity $\rho_{||}(T)$ of Nd$_3$ScBi$_5$ compared to its nonmagnetic analog La$_3$ScBi$_5$. The inset shows resistivity induced by the magnetism of Nd$^{3+}$ ions ($\Delta\rho = \rho_{Nd} - \rho_{La}$). (f) Temperature-dependent resistivity with current parallel to the c-axis at low temperatures under various magnetic fields.

To elucidate the magnetism of the frustrated kagome-lattice, we conducted anisotropic magnetic susceptibility ($\chi$) measurements (*H* =1 kOe), as shown in Fig. 1b. A distinct peak was observed at $T_{N1}$ = 5.7 K, marking the onset of the AFM order. Meanwhile, $T_{N2}$ = 4.8 K is characterized by a pronounced discontinuity, accompanied by a narrow magnetic hysteresis (~0.1 K) between zero-field-cooled (ZFC) and field-cooled (FC) measurements. At high temperatures (75-300 K), the magnetic susceptibility adheres to Curie-Weiss law, yielding an effective magnetic moment of $\mu_{eff}$ = 3.836 $\mu_B$ for *H // a* and $\mu_{eff}$ = 3.705 $\mu_B$ for *H // c*, both closely aligning with the theoretical value of 3.62 $\mu_B$ for a free Nd$^{3+}$ ion. The extracted values of $\theta_w$ = −7.082 K for *H // a* and $\theta_w$ = −64.69 K for *H // c* from the fitting represents intrinsic AFM exchange energy. It is evident that the frustration parameter $f = -\theta_w/T_{N1}$ for *H // c* ($f \approx 11.35$) is comparable to strongly geometrically frustrated magnets[12]. However, magnetic susceptibility is also significantly influenced by the excited crystalline electrical field (CEF) effect levels, as evidenced by the substantial differences in $\theta_w$ arising from pronounced magneto-crystalline anisotropy[13, 14].

A closer inspection of the magnetic ground state by AC magnetic susceptibility measurements with *H // a* elucidates the nature of the first-order phase transition. As depicted in Fig. 1c, the real component of the AC susceptibility exhibits a peak at $T_{N1}$, whereas no corresponding signal is observed in the imaginary component; notably, a discontinuous jump emerges in the real component at $T_{N2}$, accompanied by a change in the imaginary component, suggesting a FOMT to an AFM state under cooling[15]. Fig. 1d illustrates temperature-dependent magnetic susceptibility under a magnetic field of up to 80 kOe for *H // a*, demonstrating a typical AFM characteristic fingerprint. The transition at $T_{N1}$ shifts to lower temperatures with an applied field and disappears at *H* = 80 kOe, while the transition at $T_{N2}$ also moves to lower temperatures under an applied field, exhibiting a narrow hysteresis (~0.1K), and vanishes at *H* = 60 kOe. The changes in these transitions are more clearly illustrated by the *dχ/dT* analysis in Fig. S3c of Supplementary Note 3. Utilizing the Clausius-Clapeyron formalism ($dT/dH = -\Delta M/\Delta S$) to elucidate the FOMT at $T_{N2}$ reveals that increasing the magnetic field shifts the phase transition temperature towards the phase characterized by lower magnetic susceptibility.

The magnetic order in Nd$_3$ScBi$_5$ is clearly reflected in its temperature-dependent resistivity. The magnetic contribution to resistivity, denoted as $\Delta\rho = \rho_{Nd} - \rho_{La}$, exhibits a pronounced drop at $T_{N2}$ and a distinct hump around 75 K (Fig. 1e). The sharp decline reinforces the first-order nature of the magnetic phase transition, while the emergence of the hump could potentially be linked to enhanced scattering due to thermal population of excited CEF levels[19, 16]. With the application of an increasing magnetic field, the $\rho_{||}(T)$ profile becomes smoother, and the transitions shift towards lower temperatures (Fig. 1f), aligning with observations from the *χ(T)* data. From this, we deduce a profound correlation between magnetic order and electrical transport of Nd$_3$ScBi$_5$.

**Large latent heat and magnetocaloric effects**

Having established the FOMT in the frustrated kagome lattice at low temperatures, it is intriguing to explore the thermodynamic behavior of Nd$_3$ScBi$_5$. Fig. 2a presents the magnetic contribution to the specific heat $C_m$ from 150 K down to 1.8 K. An exceptionally sharp heat capacity peak at $T_{N2}$ is observed. Similar result was obtained for multiple samples from the same batch (Fig. S4a of Supplementary Note 4), confirming the first-order nature of this transition (discussed further below). Furthermore, a distinct *λ*-type anomaly, characteristic of a second-order magnetic phase transition, is observed at $T_{N1}$, along with a Schottky-like anomaly centered around 32 K, attributable to the CEF splitting of the Nd$^{3+}$ multiplet[17, 18]. The magnetic entropy $S_m(T)$, derived from integrating $C_m/T$ with respect to *T* and presented in Fig. 2b, achieves 95% of *Rln2* at $T_{N1}$ and flattens out to *Rln2*, confirming that the CEF ground state is a Kramers doublet and the excited state is significantly separated[19, 20]. The substantial value of $S_m$ near $T_{N1}$ indicates the localized character of Nd$^{3+}$ moments, and the complete recovery of the magnetic entropy (*Rln2*) occurs only around 7 K, attributable to strong short-range spin correlations[21].

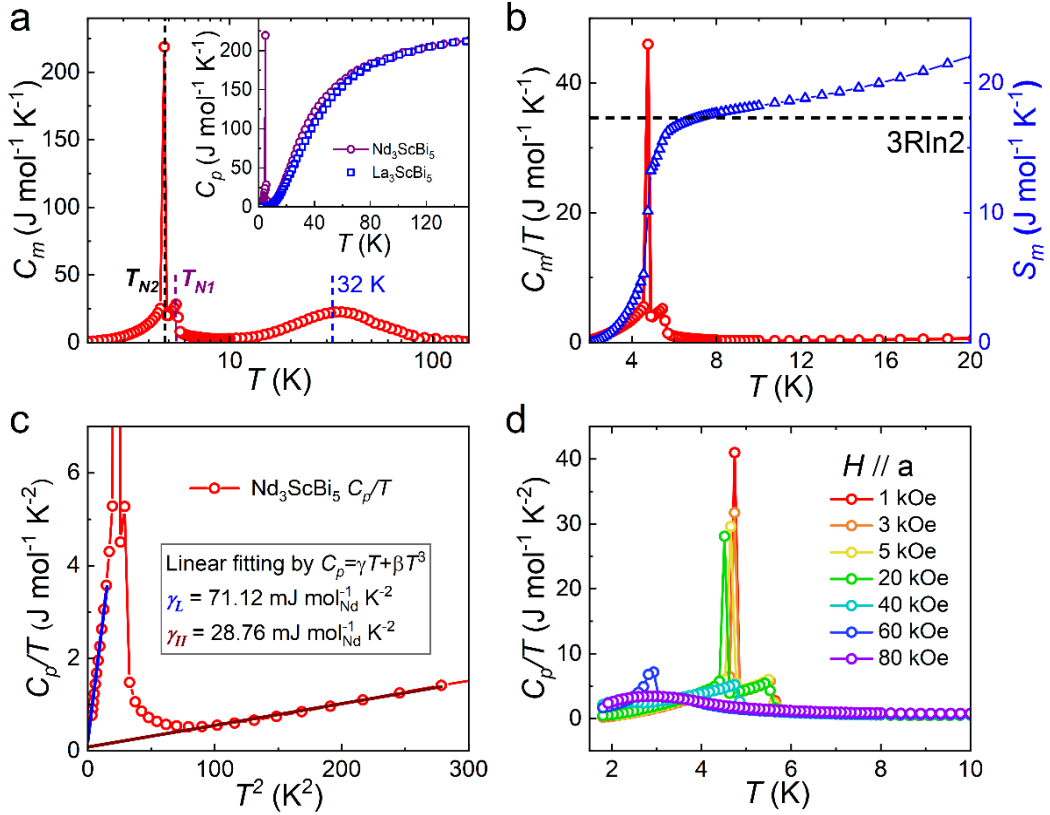

**Fig. 2. Heat capacity ($C_p$) of Nd$_3$ScBi$_5$.** (a) Magnetic contribution to the specific heat $C_m$ of Nd$_3$ScBi$_5$, with the dotted lines indicating $T_{N1}$, $T_{N2}$, and a broad peak at 32 K. The inset compares the specific heat of Nd$_3$ScBi$_5$ with that of La$_3$ScBi$_5$, where the latter reflects the combined magnetic and nuclear contributions to the specific heat of Nd$_3$ScBi$_5$. (b) Plot of $C_m/T$ versus $T$ and the corresponding magnetic entropy $S_m$. (c) $C_p/T$ versus $T$ curves of Nd$_3$ScBi$_5$ in two low-temperature windows to facilitate the estimation of the electron-specific heat coefficient $\gamma$. The blue and brown lines represent the linear fit. (d) Low-temperature $C_p/T$ versus $T$ curves under different magnetic fields applied parallel to the $a$-axis.

The Sommerfeld coefficient $\gamma_L$, estimated by linearly extrapolating $C_p/T$ vs $T^2$ from below $T_{N2}$ down to 1.8 K, stands at 71.12 mJ mol$_{Nd}^{-1}$K$^{-2}$ (Fig. 2c). The moderate $\gamma_L$ within the ordered phase provides direct insight into short-range interactions or frustration-induced quantum fluctuations[22]. Conversely, the $\gamma_H$ = 28.76 mJ mol$_{Nd}^{-1}$K$^{-2}$, obtained from fitting $C_p(T)$ in the 9-15 K temperature range above $T_{N1}$, indicates a relatively weak electron correlation effect. The application of a magnetic field along the $a$-axis progressively suppresses both transitions (Fig. 2d), as anticipated from the AFM order, exhibiting a similar field dependence to the peak in $\chi(T)$ curves.

Given the significant latent heat at the FOMT in Nd$_3$ScBi$_5$, it is reasonable to anticipate the presence of a pronounced MCE. The MCE is quantified as entropy changes, $\Delta S_M$, which can be calculated from the $M(T)$ data measured under varied applied magnetic fields (See Supplementary Note 5 for details). For field along the $a$-axis, the MCE is indeed notable, with $-\Delta S_{max}$ = 3.4 J kg$^{-1}$K$^{-1}$ at an external field of 20 kOe at $T_{N2}$ (Figs. 3a and 3b). The $M(T)$ curves measured parallel to the $c$-axis, along with the corresponding $\Delta S_M$, are presented in Fig. S6 of Supplementary Note 5. The easy-axis anisotropy of the system is evident from the significantly reduced response of $M_c(T)$ and thereby $\Delta S_{Mc}$ when compared to the measurements conducted with $H // a$, highlighting the system's potential for the rotating magneto caloric effect (RMCE)[23]. The substantial RMCE observed in the Nd$_3$ScBi$_5$ single crystal can be elucidated through the framework of single-ion magnetic anisotropy theory[24]. The results from Curie-Weiss fitting indicate that the magnetic moment of Nd$_3$ScBi$_5$ is predominantly attributed to the Nd element, implying that the magnetic anisotropy is entirely derived from the Nd$^+$ ions. Accordingly, the theory of single-ion magnetic anisotropy holds sway in magnetic materials exhibiting such local moments[25, 26]. Moreover, by computing the exponent $n$ from the power law $\Delta S_M \propto H^n$ (ref.[27]) and finding $n$ values consistently exceeding two across all fields, we further confirm the first-order nature of the $T_{N2}$ phase transition.

Heat capacity results can also be used to determine $\Delta S_M$ and adiabatic temperature changes ($\Delta T_{ad}$), but for extremely sharp transitions, these measurements invariably introduce significant errors. To mitigate this, we propose a refinement of the standard heat capacity data provided by commercial semi-adiabatic calorimeters by the point-by-point analysis of thermal response during heating and cooling in a single pulse temperature range across the FOMT (See Supplementary Note 6 for details)[28, 29]. The heat capacity peak associated with the FOMT is strikingly sharp, well exceeding the Dulong-Petit limit, with a remarkably narrow width of approximately 0.05 K at half-height, and in particular, exhibits minimal thermal hysteresis (<0.01K) observed in this system (Fig. 3c). Exhausting our current knowledge, the observation of such a profound latent heat with an almost ideal FOMT at low temperatures is rare among intermetallic compounds, except for the $Ln_2$In polycrystalline system ($Ln$ = Pr, Nd, Eu), whose FOMT occurs in the higher temperature range of 50 to 110 K[15, 30, 31]. It is well established that the heat capacity peak arises from the absorption of heat, which facilitates the randomization of magnetic moments during phase transitions, indicating the presence of a low energy barrier between the two phases preceding and following the FOMT (discussed further below)[15, 32, 33]. Figs. 3d and 3e display the indirect $\Delta S_M$ and $\Delta T_{ad}$, respectively, derived from heat capacity measurements conducted at low magnetic fields (See Supplementary Note 6 for details). The $\Delta S_M(T)$ curve displays a single peak, with the maximum value of $-\Delta S_M$ reaching 3.8 J kg$^{-1}$ K$^{-1}$ at $H$ = 20 kOe, which is comparable to the value extracted from the $M(T)$ curve. Additionally, at a magnetic field of 20 kOe, the $\Delta T_{ad}$ of Nd$_3$ScBi$_5$ attains a value of 0.9 K.

**Direct demonstration of phase transformation in Nd$_3$ScBi$_5$**
The temperature dependence of phonon frequencies at phase transition points may exhibit linear discontinuities, which can be qualitatively employed to distinguish between first- and second-order phase transitions[34, 35]. This investigation is facilitated by conducting unpolarized and polarized Raman scattering measurements using a 633 nm laser (Fig. 4a), which provides high-resolution spectra. Given the space group $P6_3/mcm$, Nd$_3$ScBi$_5$ is expected to have 10 Raman active phonon modes ($2A_{1g}+3E_{1g}+5E_{2g}$). Importantly, the $A_{1g}^2$ mode is clearly distinguishable, consistent with our symmetry analysis, and displays an exceptional signal-to-noise ratio (Supplementary Note 7). Furthermore, when comparing data taken at 1.8 K and 300 K, a pronounced red shift of the Raman peak at 300 K is evident.

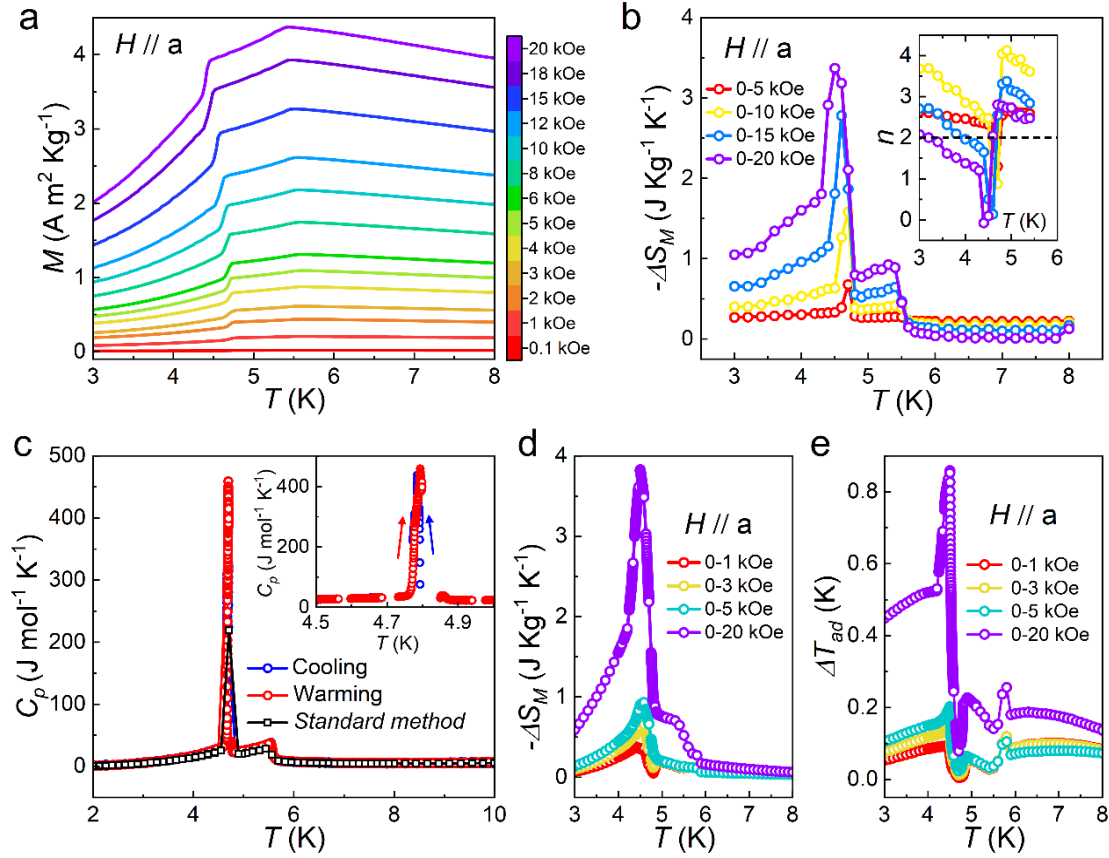

**Fig. 3. Magnetocaloric effects of Nd$_3$ScBi$_5$.** (**a**) Temperature-dependent magnetization measured upon warming under various magnetic fields. (**b**) Temperature-dependent magnetic entropy change (–$\Delta S_M$) under applied field changes determined from magnetization. The inset shows the exponent $n$ (where $|\Delta S_M| \propto H^n$) vs $T$. (**c**) Heat capacity ($C_p$) calculated using the single pulse method (SPM, circles) compared to the standard analysis of the Multiview software Quantum Design (squares). The inset displays the low-temperature closeup of $C_p(T)$ for both the warming and cooling branches with the SPM. (**d**) Temperature-dependent –$\Delta S_M$ under applied field changes derived from calorimetry. (**e**) Temperature-dependent adiabatic temperature changes ($\Delta T_{ad}$) under applied field changes computed from calorimetry.

We then investigate the Raman shift of the $A_{1g}^2$ mode phonons to assess the potential impact of Nd ordering on the phase transition of Nd$_3$ScBi$_5$. Fig. 4b displays the non-polarized Raman scattering spectrum with increasing temperatures. Initially, the peak frequency exhibits an unusual gradual increase, but after reaching 5 K, it transitions to a continuous decline toward lower frequencies. The frequency shift anomalies are likely linked to thermally induced phase transitions, which involve symmetry rearrangement and a subsequent reassignment of symmetry operations following the transition[36]. Following the damped harmonic oscillator model, a Lorentz deconvolution of the spectrum is applied, as plotted in Fig. 4c. The temperature-dependent phonon frequencies demonstrate a plateau-like softening around $T_{N1}$ and a marked discontinuity around $T_{N2}$. The anomaly observed at $T_{N1}$ can be attributed to the AFM ordering, which induces softening of the Raman modes due to strong spin-lattice coupling[37]. The sharp discontinuity near $T_{N2}$ likely arises from a discontinuous displacement of Nd ions during the phase transition, leading to lattice expansion or contraction along the interaction direction, and resulting in enhanced coupling between phonons and spins[34, 38].

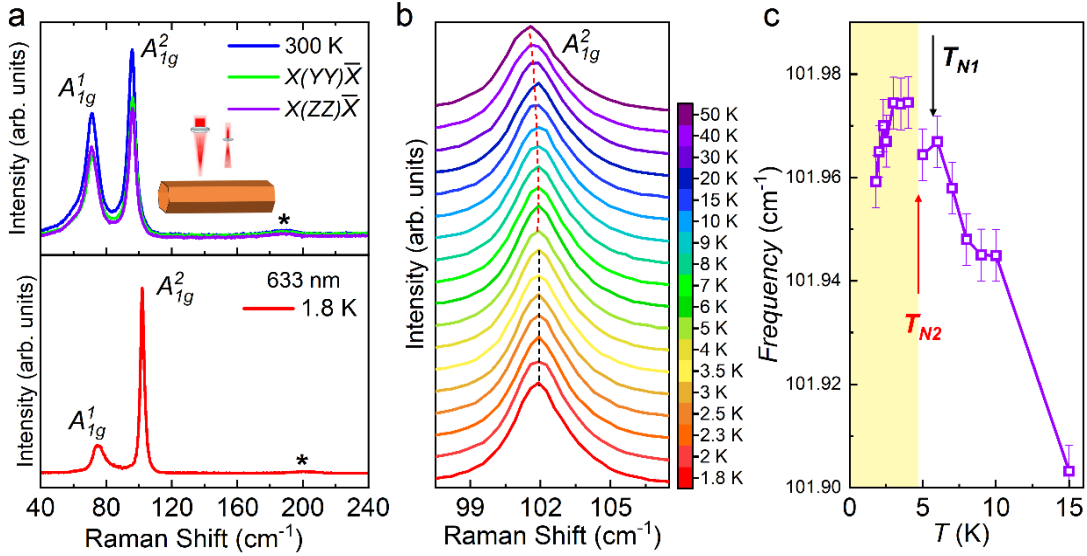

**Fig. 4. Raman scattering spectra of $Nd_3ScBi_5$.** (**a**) The Raman spectra at 300 K and 1.8 K with 633-nm lasers. Note that after determining the in-plane $a$ and collecting Raman spectra with exactly $X(YY)\bar{X}$ and $X(ZZ)\bar{X}$ configurations. (**b**) Raman spectra of the $A_{1g}^2$ mode with increasing temperatures and (**c**) temperature-dependent relative shift in the stretching mode frequency after the corresponding Lorentzian fitting.

**Quantum phases controlled by magnetic fields**
The field dependence of the magnetization at 2 K for $H$ // $a$ reveals a spin-flop and a metamagnetic transition at $H_1 = 41$ kOe and $H_2 = 66$ kOe, respectively, accompanied by slight hysteresis (Fig. 5a). Linear extrapolation of the magnetization data between $H_1$ and $H_2$ indicates that all $M$-$H$ curves converge towards the origin, as illustrated by the fit results represented by the purple line. This behavior suggests that the system remains in an antiferromagnetic state following the transition at $H_1$, substantiating the spin-flop nature of the transition[39]. Conversely, beyond $H_2$, the magnetization tends to saturate at higher fields, indicative of a metamagnetic transition. The negative slope of the Arrott diagram indicates, as per Banerjee's criterion, that the magnetic transitions in $H_1$ and $H_2$ are of the first order [Fig. S9b of Supplementary Note 8]. The field-induced transformation behavior notably aligns with the specific heat (Fig. 5a) and Raman scattering [Fig. S10 of Supplementary Note 9] data obtained under swept magnetic fields.

More interestingly, the magnetic transport data at 2 K revealed significant discontinuities at $H_1$ and $H_2$, as observed in both MR and Hall effect measurements. Combined with the above analysis of specific heat and Arrott diagram, the results show that the magnetic phase between the I and II states becomes a first-order magnetic transition, which can be induced not only by temperature but also by the applied magnetic field. In this II state, a high MR and Hall response is observed, indicating the presence of strong quantum fluctuations in this quantum critical region, suggestive of a possible gap opening[40, 41]. The comparable trends in the MR and anomalous Hall effect (AHE) during the spin-flop transition imply a shared mechanism: the transport relaxation time influencing MR and AHE is approximately proportional to the inverse of the density of states at the Fermi energy, potentially undergoing abrupt changes at the spin-flop transitions[42, 43]. After the transition at $H_2$, $Nd_3ScBi_5$ shifts from the II state to the III state (the forced-ferromagnetic (F-FM) state). It is important to note that the $Nd^{3+}$ moments are not perfectly aligned during the metamagnetic transition; as the magnetic field increases, these moments further tilt in the direction of the applied field.

It is intriguing to observe that as the temperature rises, the quantum critical behavior associated with $H_1$ and $H_2$ in the $M(H)$ curve gradually shifts towards lower magnetic fields. (Fig. S9a of Supplementary Note 8) At 5 K, the anomaly present at $H_1$ in the $M(H)$ curve vanishes, whereas the feature at $H_2$ progressively broadens and flattens, ultimately decreasing to approximately 36 kOe (Fig. 5b). Surprisingly, measurements of specific heat at 5 K indicate elevated values at low magnetic fields, providing evidence that this quantum phase is characterized by significant quantum fluctuations stemming from quantum critical behavior. Additionally, both the MR and the Hall resistivity exhibit distinct alterations in the critical field. Upon transition to the F-FM state, the MR demonstrates a linear decrease, with a sign reversal occurring at high fields, which may be associated with significant changes in the band structure near the Fermi energy caused by magnetic fluctuations[44, 45]. Upon raising the temperature to 10 K, the field-induced transformation ceases, the specific heat value at low field diminishes, and MR becomes positive (Fig. 5c).

To further elucidate the field-induced quantum phase transitions in $Nd_3ScBi_5$, magnetization measurements were conducted at 2 K for both $H // a$ and $H // c$ at high magnetic fields. (Fig. 6a) The data display linear, non-saturating behavior up to $H$ = 150 kOe, and the linear slope in both directions tends to be consistent at high fields, indicating that $Nd_3ScBi_5$ enters a field polarized regime attributed to Van Vleck (VV) paramagnetism arising from the CEF effect[46, 47]. After subtracting a linear VV paramagnetic background, the $M(H)$ curve for $H // a$ exhibits a clear fractional step-like behavior, featuring a spin-flop transition corresponding to a 1/2 magnetization plateau (Fig. 6b). At an external magnetic field of 150 kOe, the magnetization reaches only 1.05 $\mu_B$/Nd, significantly lower than the $Nd^{3+}$ saturation magnetization in vacuum (3.27 $\mu_B$), due to partial shielding of the $Nd^{3+}$ moment by spin frustration and CEF potential[21, 48, 49]. Fractional magnetization plateaus in quantum magnets signal geometrical frustration, which emerges when the spin structure becomes commensurate with the underlying lattice ordering periodicity[50]. One plausible explanation for the 1/2 magnetization plateau is a change in the ordering of local moments, which can give rise to fractional magnetization plateaus, as observed in other frustrated systems[51, 52]. Alternatively, the plateau may arise from an intermediate paramagnetic state, a phenomenon occasionally seen in certain antiferromagnetic materials[53, 54].

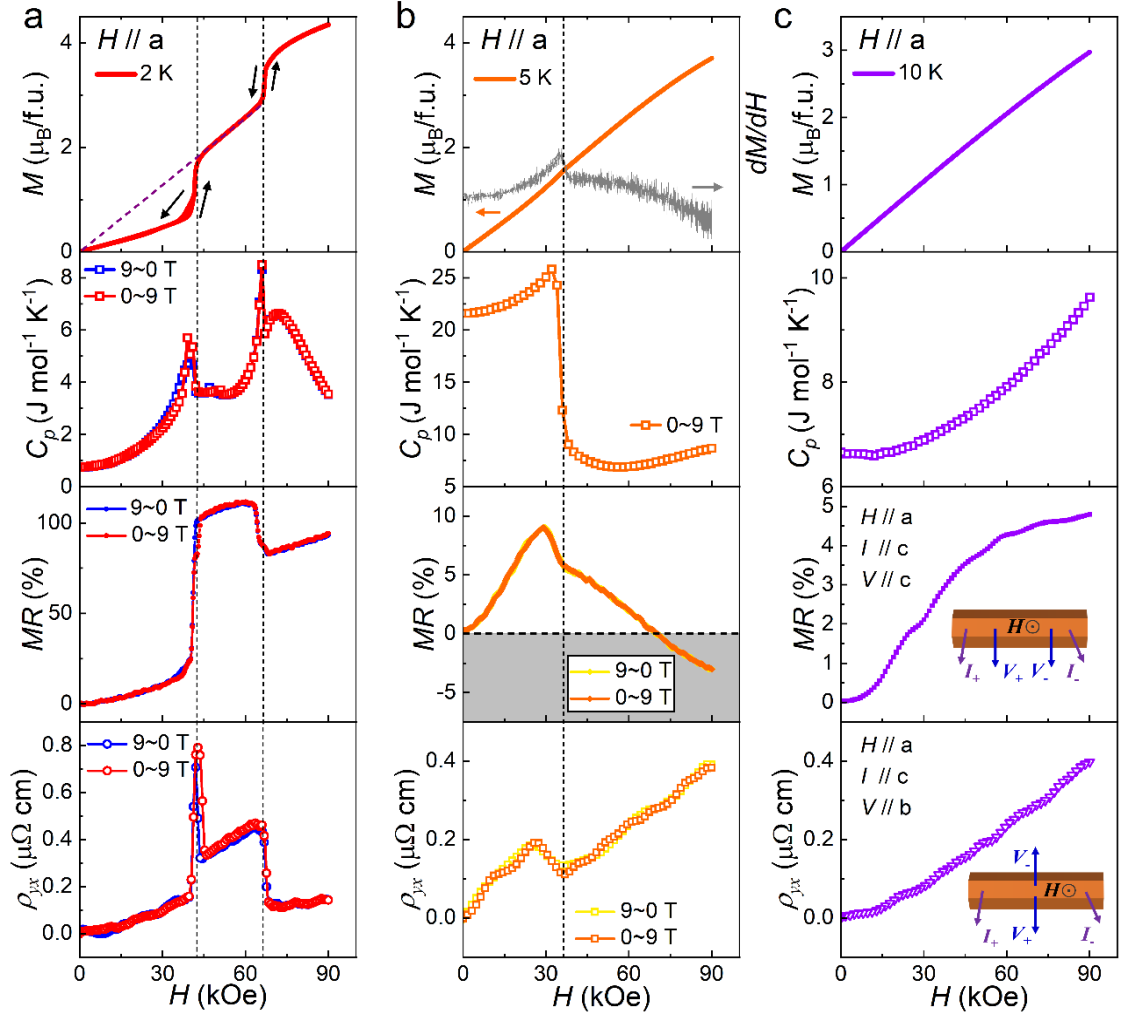

**Fig. 5. Field dependence of magnetization, specific heat, and transport properties of Nd$_3$ScBi$_5$ at low temperatures.** For $H \parallel a$, the magnetization $M$, the specific heat $C_p$, the magnetoresistance $MR$ ($I \parallel c$ and $V \parallel c$), and the Hall resistivity $\rho_{yx}$ ($I \parallel c$ and $V \parallel b$) are presented at temperatures of 2 K (**a**), 5 K (**b**), 10 K (**c**).

**Magnetic structure characteristics and magnetic phase diagram**

Through meticulous magnetic measurements performed in various crystallographic orientations, we provide deep insight into the magnetic structure of Nd$_3$ScBi$_5$. The $\chi(T)$ and $M(H)$ curves obtained for $H \parallel b$ exhibit magnitudes and trends comparable to those for $H \parallel a$ (Figs. S3a and S9c), suggesting minimal anisotropy within the plane. When aligned with the $c$-axis, Nd$_3$ScBi$_5$ demonstrates hard magnetization axis behavior, evident in its persistent magnetic order at $H = 80$ kOe and a linearly dependent magnetization profile (Figs. S3b, S3d and S9d). Integrating the aforementioned crystallographic and magnetic insights, key magnetic structure features can be summarized as **(i)** Strong coupling of $ab$ in-plane spins, leading to high magnetic moments and facilitating the rotation of Nd$^{3+}$ spins within the plane; **(ii)** Consistent two-dimensional anisotropy across the $ab$-plane; **(iii)** More restricted spin motion along the c-axis, despite shorter Nd-Nd distances in the zig-zag chain versus the kagome layer, indicating pronounced interlayer interactions[55, 56]. Higher-field magnetization and neutron-scattering measurements are needed to better elucidate the precise magnetic configuration of Nd$_3$ScBi$_5$.

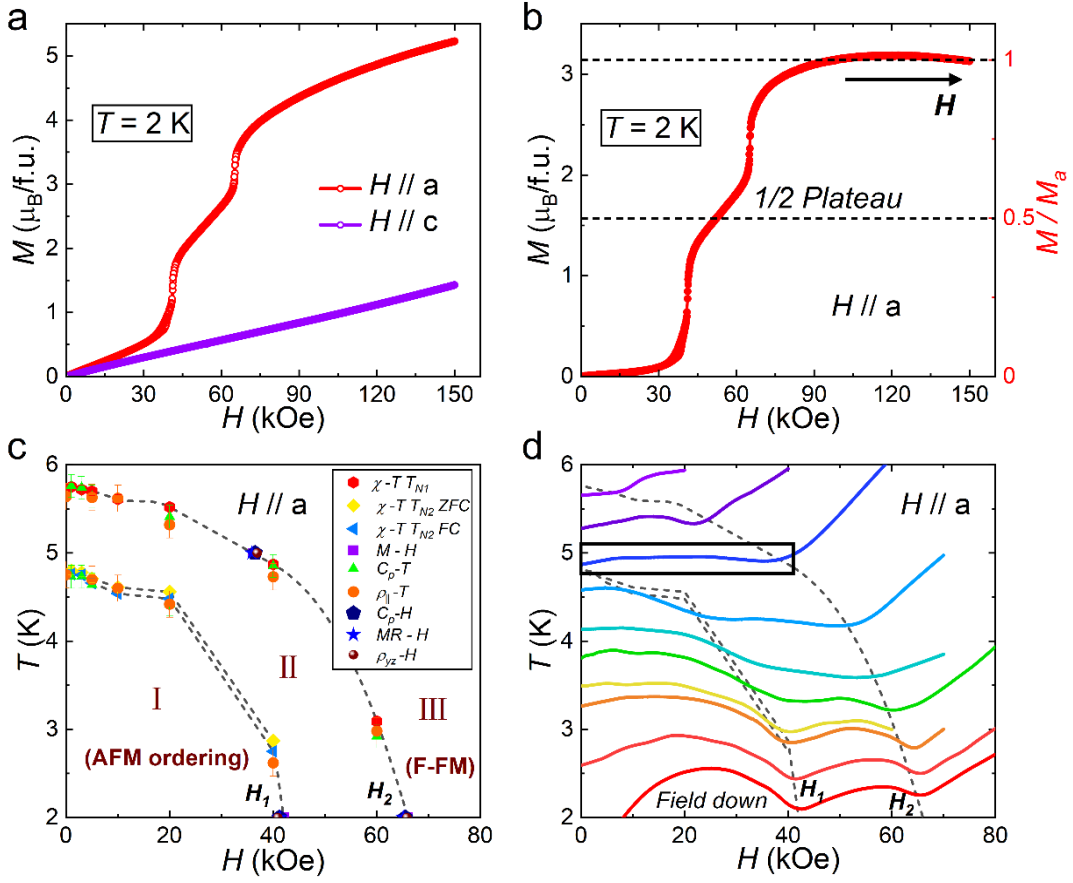

**Fig. 6. High-field magnetization, the field-temperature phase diagram, and low-temperature MCE of Nd$_3$ScBi$_5$.** (**a**) High-field dependence of magnetization at $T = 2$ K for $H \parallel a$ and $H \parallel c$ orientations. (**b**) Magnetization curves adjusted to exclude Van Vleck paramagnetism effects. (**c**) The magnetic phase diagram in the field-temperature space. The $\chi$-$T$ data points originate from sharp transitions in the $d\chi/dT$ curve. The $C_p$-$T$ and $\rho_\parallel$-$T$ data points are determined by peak positions in specific heat and resistivity measurements, respectively. Critical fields for $M$-$H$, $C_p$-$H$, $MR$-$H$, and $\rho_{yx}$-$H$ are discerned through the field dependence depicted in Fig. 5. The magnetic phase boundary (indicated by a black dashed line) is deduced from temperature-dependent susceptibility, specific heat, and resistivity curves. (**d**) Quasi-adiabatic cooling starting from various initial temperatures ($T_0 \leq 6$ K) and initial field ($B_0 \leq 80$ kOe). The phase boundaries determined from the phase diagram and the anomalies in the MCE curves are in good agreement with each other.

As depicted in Fig. 6c, collecting the critical fields and ordering temperatures determined from multiple measurements, we map the phase diagram under the $H \parallel a$ axis. There exist two critical fields: $H_2$ is nearly equivalent to the field where magnetization flats (or saturates), thus we naturally designate $H_2$ as the saturation field and $H_1$ as the critical field for the spin state transition[57]. Specifically, specific heat analysis and Arrott plot evaluation provide strong evidence that the magnetic phase transition between the I and II states exhibits a first-order nature, and it is reasonable to justify the classification of the phase boundary between the two phases belongs to the first-order transition line.

The MCE typically manifests prominently in the vicinity of field-induced magnetic phase transitions. The MCE measurements will not only facilitate the construction of phase diagrams[58], but also serve as a sensitive probe for quantum fluctuations that delineate distinct spin states. The pronounced MCE associated with fluctuating spin states underscores the potential of quantum magnets as ideal candidates for cooling applications[59]. Along each isentropic curve, adiabatic temperature changes can be read out and distinct minima are observed in the critical fields of $H_1$ and $H_2$ (Fig. 6d). The anomalous behavior of the MCE curve aligns well with the phase boundary delineated by the phase diagram, substantiating the reliability of our experimental data. It is worth noting that the isentropic line in proximity to the first-order phase transition point $T_{N2}$ presents a markedly distinct flat valley structure, in contrast to the typical isentropic bounces observed in ordered magnets, which directly reveals significant quantum spin fluctuations[3, 58]. Both the MCE and field-induced physical properties collectively suggest the presence of a strong quantum fluctuation phase in the antiferromagnetic $Nd_3ScBi_5$, characterized by enhanced MR and AHE triggered by a FOMT.

## Discussion

In summary, a novel neodymium-based intermetallic compound, $Nd_3ScBi_5$, has been unveiled to showcase the frustrated magnetism within a distorted kagome lattice. A first-order AFM transition has been identified, with large latent heat and discontinuous Raman shifts, and accompanied by minimal magnetic and thermal hysteresis. $Nd_3ScBi_5$ single crystals exhibit an extraordinarily rich and intricate magnetic phase diagram, featuring multiple distinct ground states stabilized by frustrated spin coupling and anisotropy. The field-induced response manifests a peculiar quantum critical behavior with a spin-flop and a metamagnetic transition, and reveals that the FOMT in $Nd_3ScBi_5$ can be triggered by either temperature or external magnetic fields. The significant field-sweep specific heat at low fields, coupled with the high MR and AHE states post spin-flop, indicate substantial quantum fluctuations in the vicinity of critical fields. Notably, the MCE related to a strong quantum fluctuation phase is observed in the demagnetization cooling process, manifesting itself as a prominent valley-like regime.
**Note added in proof:** As this article was about to be completed, we noticed another related work on $Nd_3ScBi_5$[60], which reported anisotropic magnetic behavior in $Nd_3ScBi_5$.

## Methods

**Sample synthesis.** High-quality single crystals of $Nd_3ScBi_5$ were synthesized via metal flux reactions with the Sc-Bi flux. The reagents included Nd (0.5785g, ≈ 4 mmol), Sc (0.3596 g, 20 mmol), and Bi (8.3592 g, 40 mmol), which were loaded into alumina crucibles and then placed into fused silica tubes with a piece of silica wool placed above the crucible for centrifugation. The tubes were flame-sealed under vacuum with an argon backfill of approximately 75 mTorr. The tubes were subjected to a temperature profile in Pit furnaces: heated to 1423 K over 10 h, maintained at this temperature for 15 h, and then cooled to 1023 K over 120 h. Upon reaching 1023 K, the tubes were removed from the furnace and inverted into a centrifuge to separate the excess flux from the rod-like $Nd_3ScBi_5$ crystals. The results presented herein are derived from the same batch of samples, though different samples prepared to ensure reproducibility exhibit comparable properties.

**Crystal structure characterization.** Single-crystal X-ray diffraction (SCXRD) of $Nd_3ScBi_5$ was performed on Kapton loops with Paratone oil on a Bruker D8 Venture diffractometer at 274 (2) K using Mo Kα radiation source (λ = 0.71073 Å). The frames were integrated with the Bruker SAINT software package using a narrow-frame algorithm. Diffraction peaks for the (h00) surface of single crystals were obtained with a Bruker D2 phaser XRD detector by using Cu Kα1 radiation (λ = 1.54184 Å). Scanning electron microscopy (SEM) and energy-dispersive X-ray (EDX) analyses were performed using a SU5000 scanning electron microscope with a Bruker EDX instrument, at an accelerating voltage of 15 kV and an accumulation time of 40 s.

**Physical property measurements.** Isothermal magnetizations and magnetic susceptibility measurements of $Nd_3ScBi_5$ conducted using a Physical Properties Measurement System (PPMS, Quantum Design, 16 T) equipped with a vibrating sample magnetometer (VSM) option and a Magnetic Properties Measurement System (MPMS, Quantum Design, 7 T). AC magnetic susceptibility measurements were conducted while warming in the MPMS under an external DC magnetic field of 10 Oe at four frequencies (13, 133, 333, and 633 Hz) after zero-field cooling. Specific heat measurements were obtained using a semi-adiabatic thermal relaxation technique (Quantum Design PPMS), with samples secured to the platform using Apiezon N grease. Longitudinal and Hall resistivities were measured simultaneously in a standard five-probe configuration using a PPMS system. To account for probe misalignment, Hall contributions to the longitudinal resistivity were corrected by averaging resistivity data obtained at positive and negative magnetic fields.

**Raman spectra measurements.** Polarized Raman spectroscopy and low-temperature Raman scattering measurements were performed on a $Nd_3ScBi_5$ single crystal with dimensions of 5 × 0.5 × 0.5 $mm^3$. The single crystal was mechanically exfoliated prior to measurements to clean the surface and reveal a fresh (100) crystallographic plane. Subsequently, the crystal was transferred into an AttoDRY 2100 cryostat, which employs helium gas as an exchange medium and enables cooling to 1.8 K. Below 100 K, the cooling rate was maintained at less than 0.5 K/min, ensuring a quasi-static cooling process. Data acquisition was carried out using a Jobin Yvon LabRAM HR Evolution spectrometer equipped with back-illuminated charge-coupled device (CCD) detector. To mitigate heating effects on the sample due to laser irradiation, a 633 nm laser with an output power of less than 10.5 μW was focused onto a spot approximately 5 μm in diameter on the sample surface.

**Quasi-adiabatic demagnetization measurements.** Isentropic curves and hold times were executed utilizing a PPMS-based setup specifically tailored for quasi-adiabatic demagnetization measurements with approximately 1.5g Sample, a system homemade and designed by Beijing National Laboratory for Condensed Matter Physics. Two pairs of twisted manganese wires, each with a diameter of 25 μm and a length of approximately 30 cm, are connected to a field-calibrated $RuO_2$ thermometer to mitigate heat leakage. Additionally, two thermal shielding plates were employed to protect the sample column from thermal radiation and other parasitic heat loads emanating from the PPMS chamber, these shields being interfaced via PEEK tubes.

# ACKNOWLEDGEMENTS

This work was supported by the National Key R&D Program of China (Grant No. 2024YFA1408400, 2021YFA1400401), the National Natural Science Foundation of China (Grant No. U22A6005, 52271238), the Center for Materials Genome, and the Synergetic Extreme Condition User Facility (SECUF, https://cstr.cn/31123.02.SECUF).

## AUTHOR CONTRIBUTIONS

Z. C. X. and X. Y. L. contributed equally to this work. Z. C. X. carried out the preparation of the samples and conducted the structure, magnetic, specific heat, and transport measurements. X. Y. L. and J. S. X. performed the quasi-adiabatic demagnetization measurements. F. J. carried out the Raman spectra analysis. C. W. Z., S. Z., and Q. S. W. participated in the discussions of the results and contributed to the manuscript preparation. X. M. Z., P. J. S. and Y. G. S. conceived the idea and supervised the project.

Supplementary Information for

# Table of Contents

# Supplementary Note 1: Single Crystal X-ray diffraction information

The structure of Nd$_3$ScBi$_5$ was determined using direct methods and refined by full matrix least-squares on $F^2$ with the *SHELXTL* program package. The refinement results are summarized in Table S1, with $R1 = 0.0195$ and $wR2 = 0.0584$ (for I > 2σ(I)) indicating a high-quality structural solution.

Table S2 provides details on the 6g, 2b, 4d, and 6g independent atomic sites for Nd, Sc, Bi1, and Bi2, along with their isotropic displacement parameters. Upon refinement, the occupancies of the Nd, Sc, and Bi Wyckoff positions were found to be fully occupied. The 6g Nd atoms are coordinated by five 6g Bi1 atoms and four 4d Bi2 atoms in a C$_{2v}$ point group, forming zig-zag chains along the *c*-axis (Fig. S1). The distance between Nd atoms within these zig-zag chains is approximately 4.9 Å, which is slightly shorter than the nearest-neighbor distance between Nd atoms in the kagome layers (~5.2 Å). The spin-exchange interactions between adjacent layers are crucial in determining the magnetic structure. Detailed anisotropic displacement parameters are listed in Table S3.

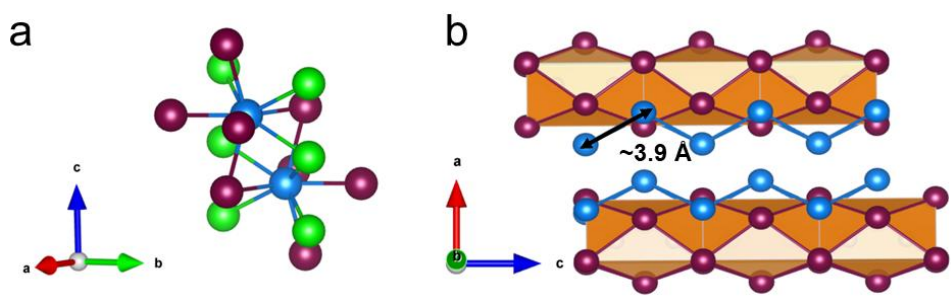

**Fig. S1. Structural features of Nd$_3$ScBi$_5$.** (**a**) The 6g Nd atoms are surrounded by five 6g Bi1 atoms and four 4d Bi2 atoms in a C$_{2v}$ point group and (**b**) forms zig-zag chains along the c axis. The blue, green, and purple spheres represent Nd, Bi1, and Bi2 atoms, respectively.

Table 1. Crystallographic and structure refinement data of $Nd_3ScBi_5$.

| Chemical formula | **$Nd_3ScBi_5$** |
|---|---|
| Temperature | 274(2) K |
| Formula weight | 1522.59 g/mol |
| Radiation | Mo $K\alpha$ 0.71073 Å |
| Crystal system | Hexagonal |
| Space group | $P6_3/mcm$ |
| Unit-cell dimensions | $a$ = 9.623(2) Å |
| | $b$ = 9.623(2) Å |
| | $c$ = 6.483(2) Å |
| Volume | 519.9(3) Å$^3$ |
| Z | 2 |
| Density (calculated) | 9.727 g/cm$^3$ |
| Absorption coefficient | 99.512 mm$^{-1}$ |
| F(000) | 1232 |
| Θ range for data collection | 2.44 to 28.28° |
| Index ranges | -12<=h<=12, |
| | -12<=k<=12, |
| | -8<=l<=8 |
| Independent reflections | 260 [$R_{(int)}$ = 0.0902] |
| Structure solution program | SHELXT 2018/2 (Sheldrick, 2018) |
| Refinement method | Full-matrix least-squares on F$^2$ |
| Refinement program | SHELXL-2018/3 (Sheldrick, 2018) |
| Function minimized | $\Sigma w\ (F_o^2 - F_c^2)^2$ |
| Data / restraints / parameters | 260 / 0 / 14 |
| Goodness-of-fit on F$^2$ | 0.778 |
| Final R indices | 249 data; I>2σ(I) |
| | R1 = 0.0195, wR2 = 0.0576 |
| | all data |
| | R1 = 0.0201, wR2 = 0.0584 |
| Weighting scheme | w=1/[σ$^2$(F$_o^2$) + (0.0599P)$^2$+1.2064P] |
| | where P=(F$_o^2$+2F$_c^2$)/3 |

Table 2. Crystallographic data of Nd$_3$ScBi$_5$.

| Atom | Wyckoff | Symmetry | x | y | z | Occup[a] | U$_{eq}$[b] |
|---|---|---|---|---|---|---|---|
| Nd | 6g | m2m | 0.0000 | 0.38227 | 0.25000 | 1.000 | 0.013 |
| Sc | 2b | -3. m | 0.00000 | 0.00000 | 0.50000 | 1.000 | 0.011 |
| Bi1 | 6g | m2m | -0.26417 | 0.00000 | 0.25000 | 1.000 | 0.012 |
| Bi2 | 4d | 3. 2 | 0.33333 | 0.66667 | 0.50000 | 1.000 | 0.012 |

[a] *Occup*: Occupancy.
[b] $U_{eq}$: equivalent isotropic thermal parameter.

Table 3. Anisotropic atomic displacement parameters (Å$^2$) for Nd$_3$ScBi$_5$.

| Label | U$_{11}$ | U$_{22}$ | U$_{33}$ | U$_{23}$ | U$_{13}$ | U$_{12}$ |
|---|---|---|---|---|---|---|
| Bi(01) | 0.0109(3) | 0.0127(3) | 0.0116(3) | 0 | 0 | 0.00637(14) |
| Bi(02) | 0.0110(3) | 0.0123(4) | 0.0162(4) | 0 | 0 | 0.00607(13) |
| Nd(03) | 0.0121(4) | 0.0121(4) | 0.0165(5) | 0 | 0 | 0.0074(2) |
| Sc(04) | 0.0113(16) | 0.0113(16) | 0.0014(3) | 0 | 0 | 0.0058(6) |

The anisotropic atomic displacement factor exponent takes the form: $-2\pi^2[ h^2 a^{*2} U_{11} + ... + 2hka^*b^*U_{12}]$

Supplementary Note 2: Sample structure characterization

As depicted in Fig. S2a, the grown crystal exhibits rod-like morphology, with the X-ray diffraction (XRD) pattern of single crystal revealing that the natural surface corresponds to the ($h$00) plane, which is the primary focus of this study due to its significant properties. We conducted energy-dispersive X-ray (EDX) analysis on multiple crystals of $Nd_3ScBi_5$ from the same batch, where the growth composition ratio was Nd: Sc: Bi = 1: 5: 10. Prior to measurement, the crystals were cleaved to expose clean surfaces.

The scanning electron microscope (SEM) images of $Nd_3ScBi_5$, corroborate the XRD results, as no secondary phases are detected in any of the samples (Fig. S2b). EDS mapping results (Figs. S2c and S2d) indicate that the elements within the $Nd_3ScBi_5$ compounds are uniformly distributed, demonstrating their high crystalline quality.

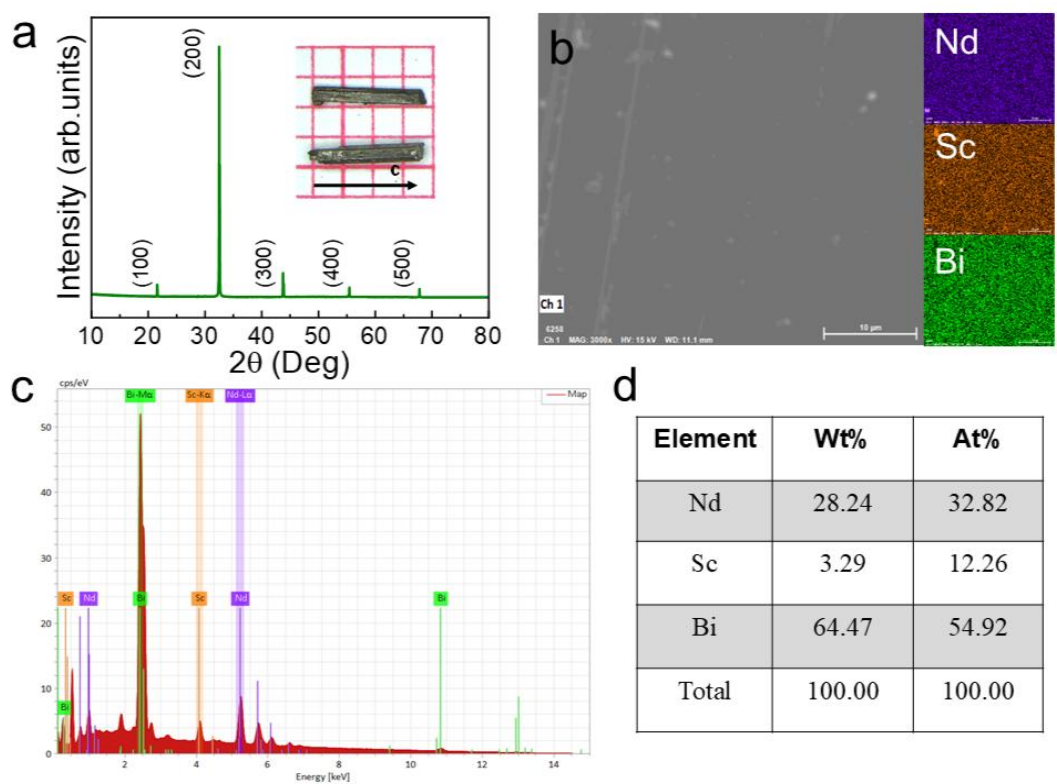

**Fig. S2. Characterization of $Nd_3ScBi_5$.** (**a**) Single-crystal X-ray diffraction (XRD) pattern recorded for the (h,0,0) planes at room temperature. The inset presents a typical image of the grown crystal on a 1-mm grid paper. The scanning electron microscope (SEM) image for single crystal and energy-dispersive X-ray (EDX) elemental color mapping for Nd, Sc, and Bi for the area in (**b**) with corresponding EDX spectra in (**c**). (**d**) weight and atomic percentage of Nd, Sc, and Bi atoms.

# Supplementary Note 3: Magnetic susceptibility and its derivatives

The magnetic susceptibility data under $H \,/\!/\, b$ shown in Fig. S3a exhibit a similar magnitude and trend as those under $H \,/\!/\, a$, with the exception of the absence of the transition at $T_{N2}$ when $H = 40$ kOe. Given the minimal in-plane anisotropy of magnetic susceptibility, we focus our systematic study of Nd$_3$ScBi$_5$ on the direction for $H \,/\!/\, a$, neglecting the anisotropy within the $ab$ plane.

Moreover, with the increase of the external field along the $c$-axis, both transitions at $T_{N1}$ and $T_{N2}$ shift slightly towards lower temperatures (Fig. S3b), a behavior that persists even at $H = 80$ kOe. These changes in the transitions for $H \,/\!/\, a$ and $H \,/\!/\, c$ are more clearly demonstrated by the derivative $d\chi/dT$ (Figs. S3c and S3d).

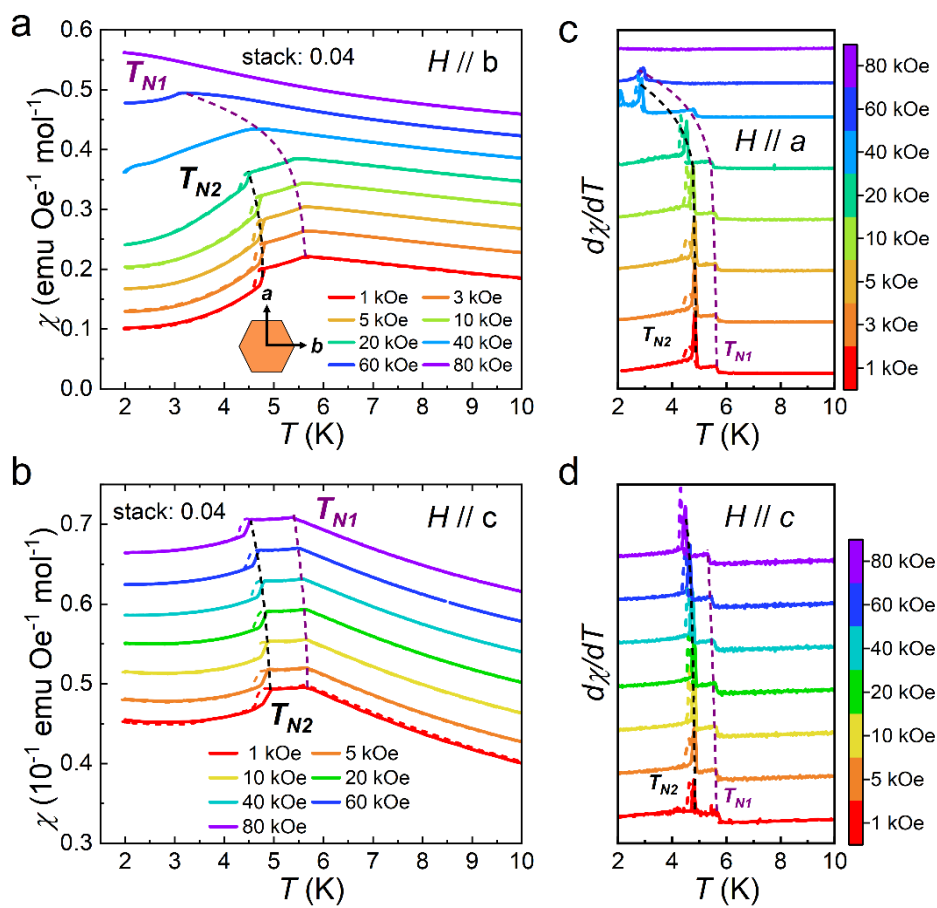

**Fig. S3. Supplemental magnetic susceptibility data of Nd$_3$ScBi$_5$.** (**a, b**) Temperature-dependent magnetic susceptibility under different magnetic fields for $H \,/\!/\, b$ and $H \,/\!/\, c$, with zero-field cooling (ZFC) and field cooling (FC) data shown by solid and dashed lines, respectively. (**c, d**) Derivative of magnetic susceptibility ($d\chi/dT$) for Nd$_3$ScBi$_5$ under various magnetic fields for $H \,/\!/\, a$ and $H \,/\!/\, c$. Data have been shifted vertically for clarity, and the short dot lines are guided to the eye.

# Supplementary Note 4: Supplemental specific heat data

Due to the challenges in accurately measuring the mass of these highly air-sensitive samples, there is some variability in the overall scale of the heat capacity. Additionally, slight sample-to-sample variations in the heat capacity are observed. However, the three samples of $Nd_3ScBi_5$ are generally consistent, particularly in terms of the positions of the phase transition points and the trend of the specific heat $C_p(T)$ curve. It is important to note that all other specific heat data presented in this paper are derived from $Nd_3ScBi_5$ Sample 1.

Furthermore, low-temperature $C_p(T)$ versus $T$ curves for $H$ // c are shown in Fig. S4b. It is evident that for $H$ // c, the peaks shift towards lower temperatures with increasing magnetic fields, consistent with the behavior observed in the $\chi(T)$ curve under applied fields (Figs. S3b and S3d). It should also be noted that the single crystal was detached from the sample puck during the $C_p$ measurement when the field exceeded 40 kOe.

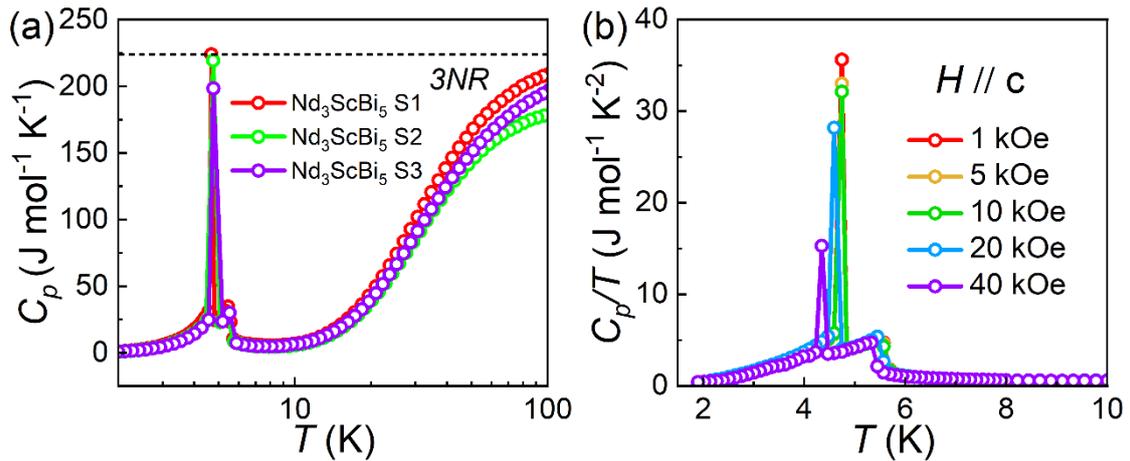

**Fig. S4. Supplemental specific heat data of $Nd_3ScBi_5$.** (a) Temperature dependence of the heat capacity for several samples. The dashed horizontal line corresponds to the Dulong-Petit limit value *3NR* at high temperatures, where *N* is the number of atoms per formula and *R* is the ideal gas constant. (b) The low-temperature $C_p/T$ versus $T$ curves with varied magnetic fields for $H$ // c.

# Supplementary Note 5: Magnetization measurements to characterize magnetocaloric effects

The most common method for measuring magnetic entropy change $\Delta S_M$ involves plotting the temperature and magnetic field dependence of the sample $M(H, T)$, then the Maxwell relationship can be used to calculate $\Delta S_M$ as follows:

$$\Delta S_M(T, H) = \int_0^{H_{max}} \left(\frac{\partial M}{\partial T}\right)_H dH \qquad (1)$$

The former method is to integrate the area between the two magnetized isotherms to obtain $\Delta S_M$ at the intermediate temperature, while the latter method is that $\partial M/\partial T$ can be found directly and $\Delta S_M$ can be found by numerical integration.

Fig. S5 illustrates the hysteretic transition of the first-order material between the low-temperature antiferromagnetic (AFM) state and the high-temperature ferromagnetic (PM) state. During isothermal magnetization tests, if the temperature is within the range $T_{low}$ to $T_{high}$, and no another phase boundary is crossed while applying the field, the transition from AFM to PM states remains incomplete. Consequently, subsequent measurements may show a coexistence of phases, resulting in a series of plateau features in the isothermal magnetization curves[1]. This issue can be avoided by performing isofield magnetization measurements where the sample is initially in a fully AFM or fully PM state. Therefore, for first-order materials, using isofield measurements of magnetization is considered more precise and accurate for characterizing the magnetocaloric effects[2].

Compared with the $H // a$ direction, the magnetocaloric contribution along the $H // c$ direction is significantly smaller. The isofield magnetization $M_c(T)$ curves of $Nd_3ScBi_5$ under applied fields up to 20 kOe are shown in Fig. S6a. Due to the magnetic anisotropy of $Nd_3ScBi_5$, the magnetization along $H // c$ is considerably lower than that along $H // a$; however, the overall trend of the $M(T)$ curves remains consistent. Using the Maxwell relation, the magnetic entropy change at low fields for $H // c$ is determined (Fig. S6b). A distinct step-like feature at the peak of the $\Delta S_{Mc}(T)$ curve indicates a first-order phase transition. The max values of $-\Delta S_{Mc}$ are 0.02, 0.08, 0.17, and 0.30 J mol$^{-1}$ K$^{-1}$ at 5, 10, 15, and 20 kOe, respectively, which are notably smaller compared to those observed in the $H // a$ direction.

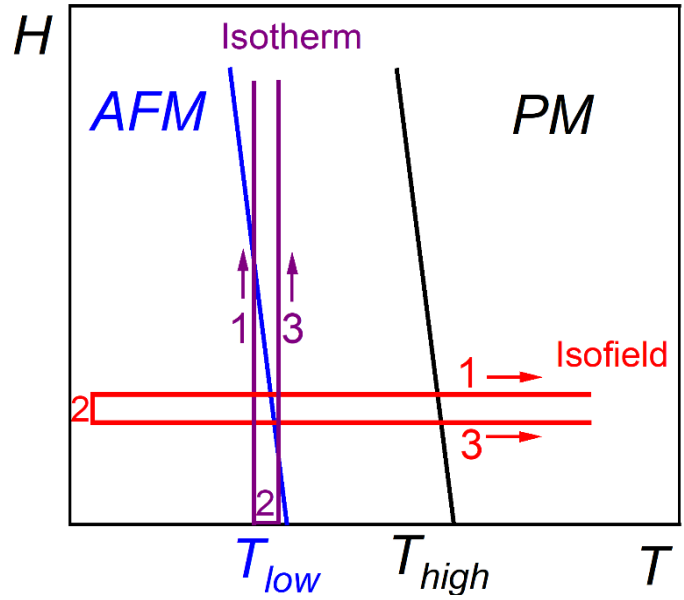

**Fig. S5. Schematic phase diagram of the antiferromagnetic first-order phase transition material.** The arrow represents the direction of measurement. The purple path (1-2-3) represents the isothermal magnetization measurement, and the red path (1-2-3) represents the isofield magnetization measurement.

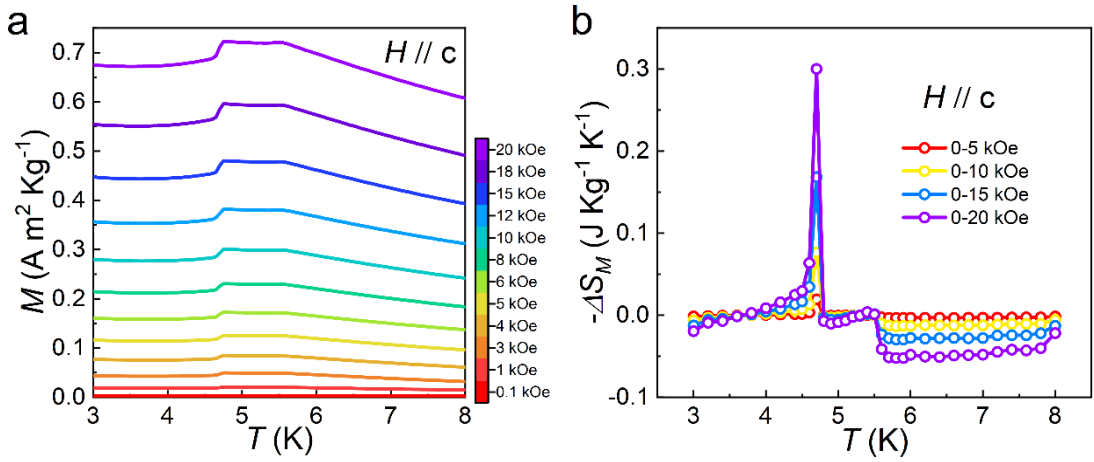

**Fig. S6. Magnetocaloric effects of $Nd_3ScBi_5$ for $H // c$.** (**a**) Temperature-dependent magnetization measured upon warming under various magnetic field. (**b**) Temperature-dependent magnetic entropy change ($-\Delta S_M$) under applied field changes determined from magnetization.

# Supplementary Note 6: Single pulse method for heat capacity measurements

Classical heat capacity measurements, using a non-zero heat capacity adiabatic pulse calorimeter, rely on a global analysis of the thermal response to small heat pulses. However, this method introduces inherent errors around first-order phase transitions[3]. These errors, when propagated to calculations of magnetocaloric effects, result in significant systematic inaccuracies[4]. For instance, in the case of $Nd_3ScBi_5$ near $T_{NI}$ (Fig. S7), the thermal relaxation curve, which crosses the first-order transition line, often lacks critical detail due to the limitations of classical curve fitting, particularly around the latent heat release. To overcome this limitation, we employed a single pulse method (SPM) that, by analyzing the point-by-point thermal response during both heating and cooling within the temperature range of a single pulse, achieves higher temperature resolution, as illustrated by the arrow trends in the Fig. S7. This approach allows for a more accurate depiction of heat capacity changes and avoids underestimating narrow features in $C_p(T)$.

The heat capacity curves presented in Figs. 3c-3e were obtained using single pulse data for heating in the 4-5 K temperature range, rather than the conventional Quantum Design software analysis. The $\Delta S_M$ can also be calculated from $C_p$ data under the zero magnetic field and applied magnetic field by using the equation:

$$-\Delta S_M(T,H) = -\int_0^T \frac{C(H,T) - C(H_0,T)}{T} dT = S(H_0,T) - S(H,T) \quad (2)$$

Fig. S8 presents the total entropy curves constructed from the heat capacity measurements. The isothermal entropy change represents the entropy variation with changes in the external magnetic field under isothermal conditions. Adiabatic temperature change $\Delta T_{ad}$, which quantifies the intuitive magnetocaloric response, can be obtained from the following equation:

$$\Delta T_{ad} = -\int_0^{H_{max}} \frac{T}{C_p} \left(\frac{\partial M}{\partial T}\right)_H dH \approx -\frac{T \Delta S_M}{C_p} \quad (3)$$

The validity of this approximation is supported by the typical shape of the $C_p(T)$ curves at different applied magnetic fields. For temperatures far from the transition temperature, the influence of the magnetic field on $C_p$ is negligible, thereby justifying the applicability of Equation 3 in these cases[5].

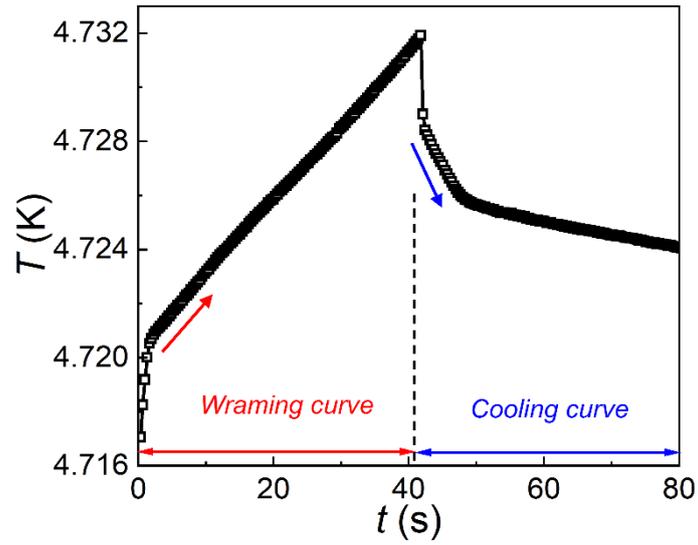

**Fig. S7.** Heat capacity relaxation. Temperature (*T*) versus time (*t*) relaxation curves of a single pule of FOMT near $T_{N1}$ in Nd$_3$ScBi$_5$ at zero magnetic field. The curves represent both heating and cooling branches. Arrows indicate the transition points upon warming (red) and cooling (blue).

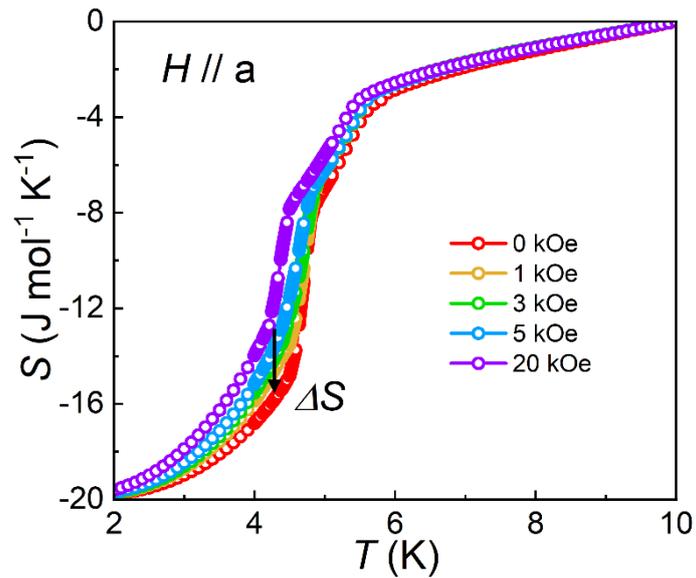

**Fig. S8.** Total entropy constructed from heat capacity for *H* // a in magnetic fields of 0, 1, 3, 5, 20 kOe.

## Supplementary Note 7: Raman Spectra Analysis

The single crystal of $Nd_3ScBi_5$, which belongs to space group 193 ($P6_3/mcm$) and point group $D_{6h}$ (6/mmm), possesses Raman tensors for the $A_{1g}$, $E_{1g}$, and $E_{2g}$ modes with the following forms:

$$A_{1g} = \begin{pmatrix} a & 0 & 0 \\ 0 & a & 0 \\ 0 & 0 & b \end{pmatrix}, \quad E_{1g} = \begin{pmatrix} 0 & 0 & 0 \\ 0 & 0 & c \\ 0 & c & 0 \end{pmatrix}, \begin{pmatrix} 0 & 0 & -c \\ 0 & 0 & 0 \\ -c & 0 & 0 \end{pmatrix},$$

$$E_{2g} = \begin{pmatrix} d & 0 & 0 \\ 0 & -d & 0 \\ 0 & 0 & 0 \end{pmatrix}, \begin{pmatrix} 0 & -d & 0 \\ -d & 0 & 0 \\ 0 & 0 & 0 \end{pmatrix}.$$

note that after determining the in-plane $a$, the Raman intensity of the $A_{1g}$, $E_{1g}$, and $E_{2g}$ modes are as follows[6]:

$$I_{A1g} \propto (a\cos^2\theta + b\sin^2\theta)^2; \quad I_{E1g} \propto c^2(4\sin^2\theta\cos^2\theta + 1); \quad I_{E2g} \propto 2d^2\cos^4\theta.$$

In the context of $(X(YY)\bar{X})$ configurations, where the $\theta$ is set to 0°, the Raman tensors dictate that $A_{1g}$ modes are suppressed in the cross configuration. The $E_{1g}$ mode remains non-extinct in both parallel and cross configurations, whereas the $E_{2g}$ mode is absent in the parallel polarization configuration $(X(ZZ)\bar{X})$.

In conjunction with the experimental findings presented in Fig. 4a of the main text, the scattering peak at 188.737 cm$^{-1}$ exhibits relatively weak intensity, necessitating further analysis and discussion. In contrast to the non-polarized scenario, the peaks at 71.375 cm$^{-1}$ and 95.485 cm$^{-1}$ persists under the parallel configurations of $(X(YY)\bar{X})$ and $(X(ZZ)\bar{X})$, thereby eliminating the possibility of it being the $E_{2g}$ mode. In subsequent testing, supplementary experiments were carried out in the cross-configuration, where both peaks showed significant suppression but were not entirely extinguished. This suggests that these two peaks might belong to the $A_{1g}$ mode. This discrepancy between theoretical expectations and experimental outcomes could be influenced by two primary factors. First, the incomplete extinction observed might arise from experimental inaccuracies, including imprecise laser alignment. Additionally, the surface of the test sample is not completely flat, which could inadvertently introduce contributions from extraneous surface signals.

## Supplementary Note 8: Magnetization and Arrott diagram

The Arrott diagram is a valuable tool for determining whether a phase transition is first-order or second-order through magnetic measurements. In the presence of an external magnetic field, the free energy ($F$) can be expressed as a function of the magnetization ($M$) (the order parameter) and the external magnetic field ($H$), as illustrated in Equation 4:

$$F(M,H) = \frac{a}{2}M^2 + \frac{b}{4}M^4 - MH \qquad (4)$$

Under equilibrium conditions, where the free energy is minimal ($\partial F/\partial M = 0$), the aforementioned equation can be simplified to Equation 5:

$$\frac{H}{M} = a + bM^2 \qquad (5)$$

In this case, the order of the phase transition is determined by the sign of the parameter $b$ [7,8], as shown in Fig. S9b, where the apparent negative slope at 2 K indicates that the field-induced magnetic phase in La$_3$ScBi$_5$ become a first-order phase transition.

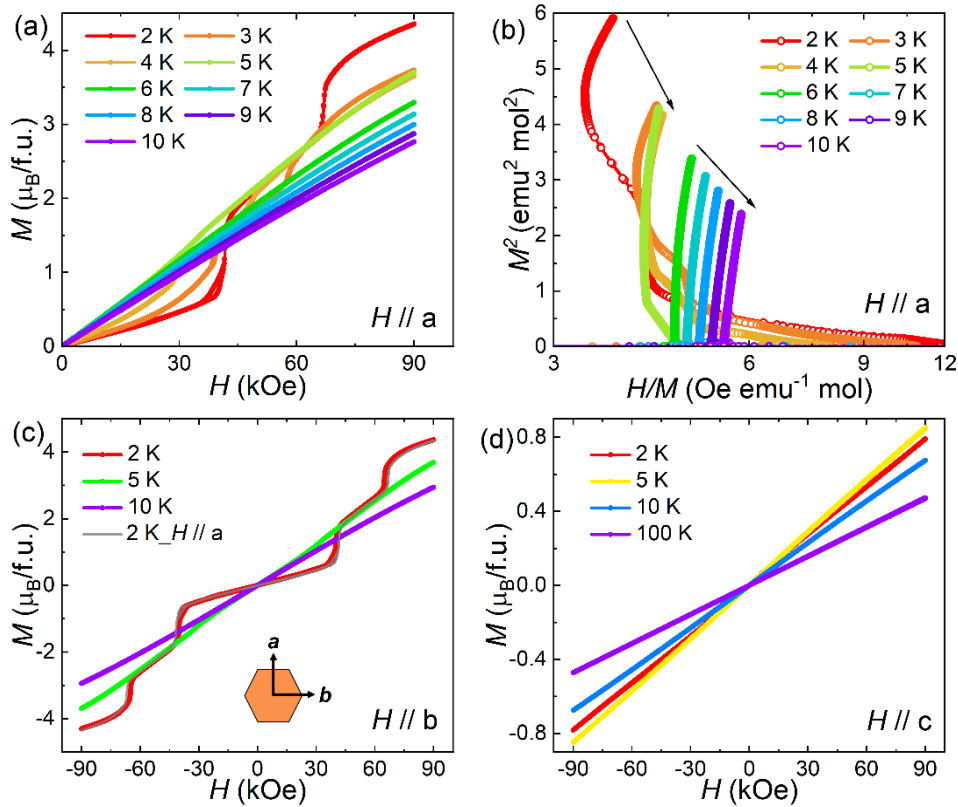

**Fig. S9. Magnetization and Arrott diagram of Nd$_3$ScBi$_5$.** (a) Isothermal magnetizations curves for various magnetic fields at temperatures between 2 K and 10 K for $H \mathbin{/\mkern-5mu/} a$. (b) Arrott plots $M^2$ vs $H/M$ for $H \mathbin{/\mkern-5mu/} a$. (c) The isothermal magnetization of $H \mathbin{/\mkern-5mu/} b$ under different magnetic fields, particularly evident in the $M(H)$ curves at 2 K. (d) The isothermal magnetization for $H \mathbin{/\mkern-5mu/} c$ under varying magnetic fields exhibits a linear field dependence, characteristic of antiferromagnetic behavior.

## Supplementary Note 9: Magnetic field dependent Raman scattering

As illustrated in Fig. S10, the $E_{1g}$ mode is prominently evident in the $H = 0$ kOe spectrum at $E_{1g} = 101.929$ cm$^{-1}$ for $H // a$. As the magnetic field increases, the frequency exhibits a sharp decline between 20 kOe and 45 kOe. Additionally, a discontinuous transition may occur around 65 kOe. These two abrupt decreases may suggest the presence of first-order phase transitions, indicating the occurrence of two spin flop-like transitions as the external field increases. Collectively, the two transitions correspond to the critical points $H_1$ and $H_2$ in the $M(H)$ curve at 2 K. They may manifest either as discontinuous jumps or as a continuous rotation of the spins.

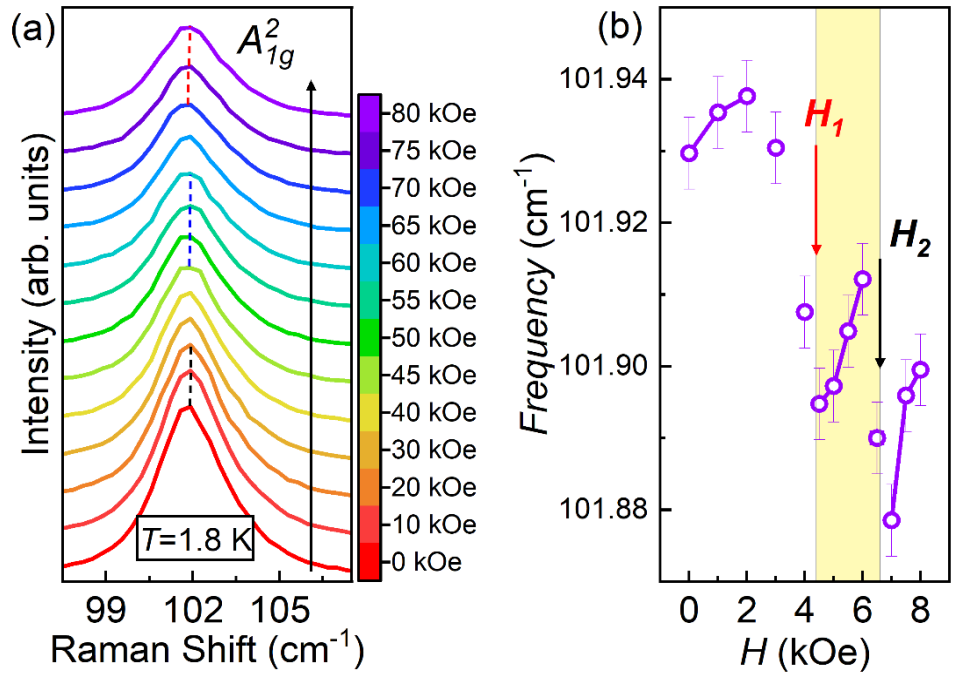

**Fig. S10. The $E_{1g}$ phonons from the magnetic field-dependent Raman scattering spectra of Nd$_3$ScBi$_5$ at 1.8 K, with the external field oriented parallel to the a-axis.** (**a**) Raman spectra as a function of increasing magnetic field. (**b**) Field-dependent relative shifts in the stretching mode frequency following Lorentzian fitting.

## Supplementary Note 10: Supplementary magnetocaloric data and phase diagrams

As depicted in Fig. S11, under diverse initial conditions near the phase transition points ($T$ = 5 K, $H$ = 70 kOe and $T$ = 6 K, $H$ = 60 kOe), the maximum adiabatic temperature change approaches 1 K, closely aligning with the values derived from specific heat measurements. The sample experiences gradual heating due to heat leakage. For instance, at an ambient temperature of 5 K, the device remains below 4.3 K for 900 minutes post-field elimination. The minimum attainable temperature and the duration of stability likewise affirm the approximate adiabatic conditions achieved by the device. Consequently, $Nd_3ScBi_5$ demonstrates commendable cooling performance.

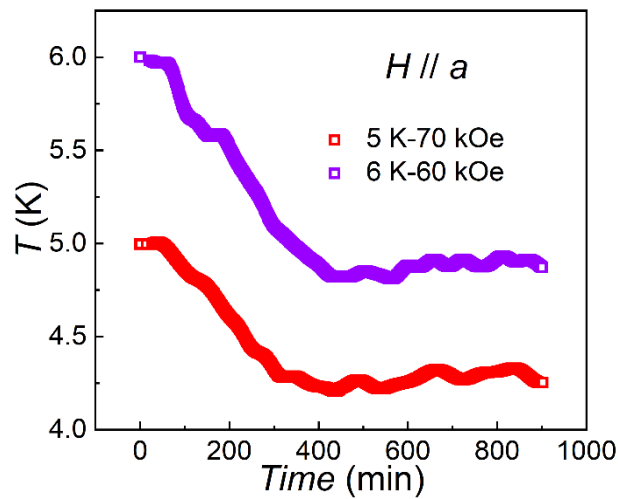

**Fig. S11.** Holding times of $Nd_3ScBi_5$ for $H$ // a at the lowest achievable temperature under different initial conditions after demagnetization.

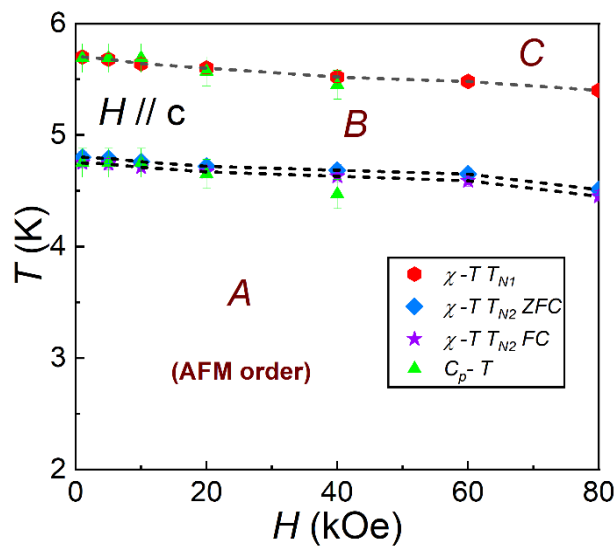

**Fig. S12.** Temperature-magnetic field phase diagram of $Nd_3ScBi_5$ for $H$ // c.

# Supplementary References: